\documentclass[onecolumn]{article}
\usepackage[utf8]{inputenc}
\usepackage[margin=1in]{geometry}

\title{Species Tree Estimation Using ASTRAL: Practical Considerations}
\author{Siavash Mirarab}
\date{}

\usepackage{natbib}
\usepackage{graphicx}
\usepackage{url}
\usepackage{amsmath,amsthm}
\usepackage{xcolor}
\usepackage{xstring}
\usepackage{xfrac}
\usepackage{soul}

\usepackage[font=small,labelfont=bf]{caption}

\newcommand{\FILL}[1]{}

\newtheoremstyle{probs}
{}
{}
{\em}
{}
{\bf}
{:}%
{ }%
{\thmnote{ #3}}
\theoremstyle{probs}
\newtheorem{problem}{Problem}

\DeclareMathOperator*{\argmax}{arg\,max}

\newcommand{\LS}{\mathcal{L}} 
\newcommand{\lset}{\LS}
\newcommand{\GT}{\mathcal{G}} 
\newcommand{\gset}{\GT}
\newcommand{\TR}{T} 
\newcommand{\T}{T}
\newcommand{\ST}{T^*} 
\newcommand{\qs}[1]{\mathcal{Q}(#1)} 

\newcommand{\X}{\mathcal{X}}

\newcommand{\fqs}[1]{
\ifx&#1&%
\bar{z}%
\else%
z_{#1}%
\fi%
}

\newcommand{\inters}[2]{I({#2,#1})}
\newcommand{\oth}[2]{h_{#1,#2}}

\renewcommand{\setminus}{-}

\begin{document}

\maketitle

\begin{abstract}
ASTRAL is a method for reconstructing species trees after inferring a set of gene trees and is increasingly used in phylogenomic analyses. 
It is statistically consistent under the multi-species coalescent model, is scalable, and has shown high accuracy in simulated and empirical studies. 
This chapter discusses practical considerations in using ASTRAL, 
starting with a review of published results and pointing to the strengths and weaknesses of species tree estimation using ASTRAL. 
It then continues to detail the best ways to prepare input gene trees, interpret ASTRAL outputs, and perform follow-up analyses. 
\end{abstract}

\section{Introduction}

\begin{figure}[b!]
\centering
\includegraphics[width=.94\textwidth]{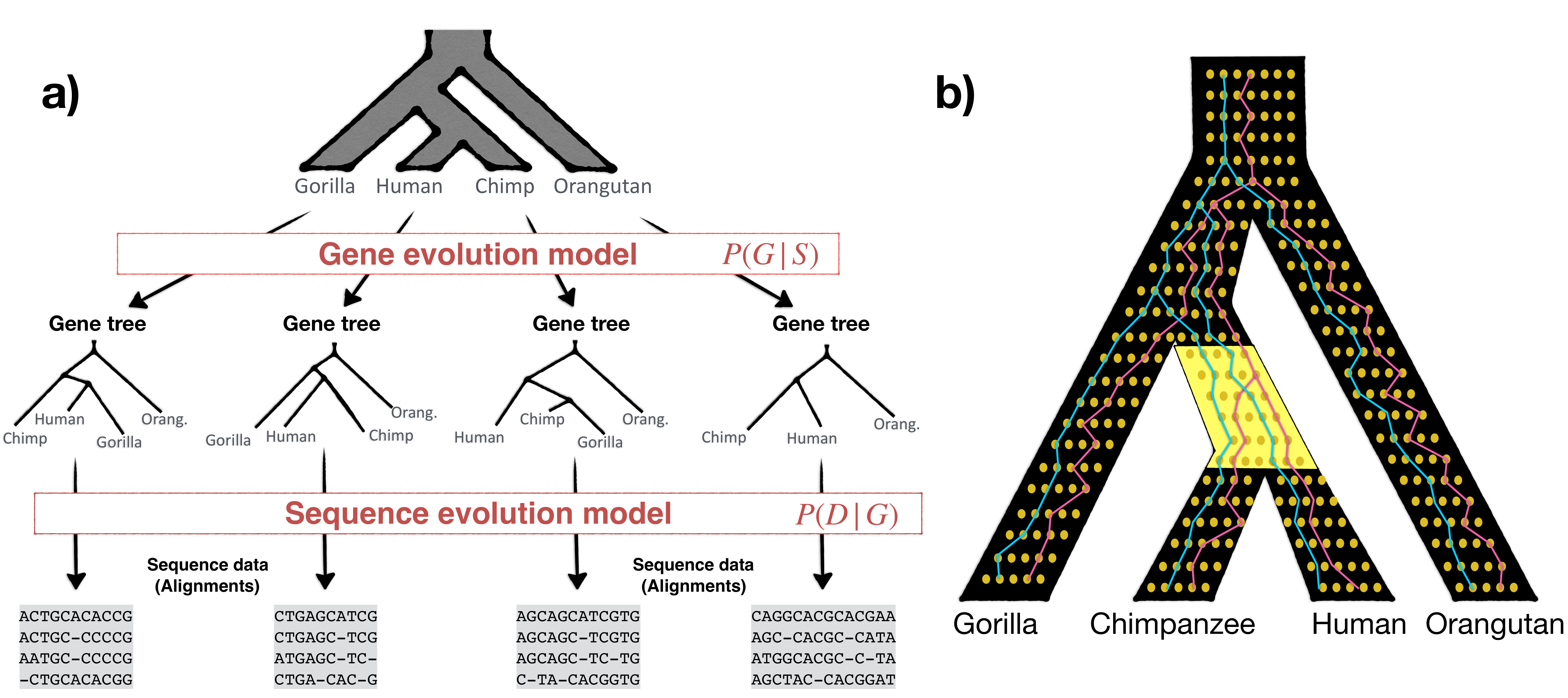}
\caption{(a) The hierarchical model of genome evolution: the species tree parameterizes a model of gene tree evolution; gene trees, sampled from this model, parametrize a model of sequence evolution, which generates the sequences. 
(b) Tracing two lineages inside a species tree where each branch is a population.
Pink lineages coalesce in ways that match the species tree topology. 
Cyan lineages fail to coalesce in the common ancestor of Human and Chimpanzee (yellow population), giving the cyan lineage from Chimpanzee a chance to coalesce with Gorilla before coalescing with Human -- and creating Incomplete Lineage Sorting (ILS). }
\label{fig:model}
\end{figure}

Understanding gene trees as entities evolving within species trees, the framework nicely summarized by \cite{Maddison1997}, has given statisticians a powerful model to approach genome-wide phylogenetic reconstruction. 
Genome evolution can be understood using a hierarchical generative model (Fig.~\ref{fig:model}a):
gene trees are first sampled from a distribution defined by a model of gene evolution and parameterized by the species tree; then, sequences are sampled from distributions defined by a model of sequence evolution and parameterized by the gene trees and other necessary parameters. 
The choice of the exact model of sequence evolution
and the model of gene tree evolution defines the exact hierarchical model.

A leading model of gene evolution is the multi-species coalescent (MSC) \cite[see][]{Degnan2009}. 
\FILL{and Chapters \ref{sec:chapter?} of this book.} 
MSC models incomplete lineage sorting (ILS) and the resulting discordance between gene trees and the species tree 
(Fig.~\ref{fig:model}b). 
Note that I use terms \textit{gene} and \textit{locus} interchangeably to refer to a recombination-free region of the genome (not functional genes). 
The MSC model is widely adopted due to its perceived biological realism and  mathematical convenience. 
Under MSC, the species tree is identifiable from a distribution of gene trees~\citep{Allman2011}, giving us hope to recover the species tree from gene trees. 

\begin{figure}[t]
\centering
\includegraphics[width=0.9\textwidth]{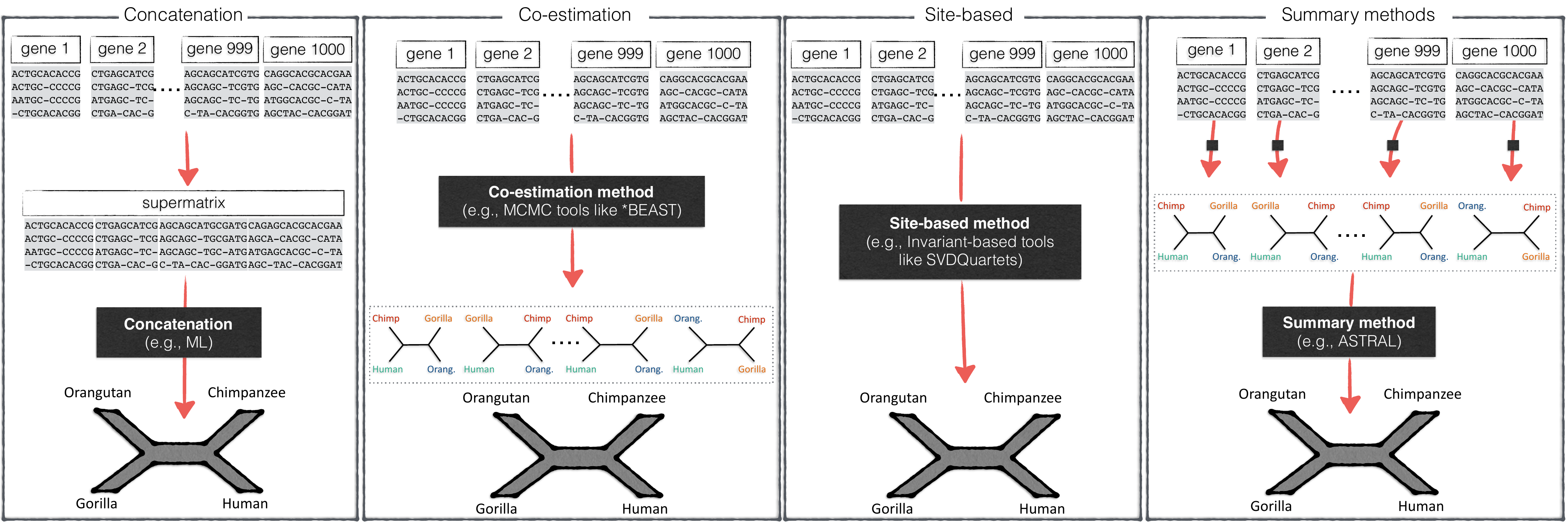}
\caption{Four main approaches to species tree estimation. }
\label{fig:methods}
\end{figure}

Several approaches exist for inferring the species tree given multi-gene sequence data under MSC. 
Concatenating sequences from all loci and performing ML inference under a model of sequence evolution (Fig.~\ref{fig:methods}) amounts to ignoring the gene evolution component of the hierarchical model (Fig.~\ref{fig:model}) and is proved by \cite{Roch2014} not to be statistically consistent.
This inconsistency, predicted earlier by \cite{KubatkoDegnan2007}, has motivated the development of alternative MSC-based approaches. 

Given the hierarchical nature of the model, the most statistically principled approach
is to co-estimate gene trees and the species trees as part of one joint inference (Fig.~\ref{fig:methods}). 
Methods of co-estimation  have been developed, mostly using Bayesian MCMC to sample the distributions defined by the hierarchical model \citep[e.g.,][]{best,Heled2010} and have been shown in simulations to have good accuracy under the MSC model \citep{Bayzid2013,Ogilvie2016}.
These methods, however, need to sample a vast number of parameters: topologies of the species tree and all gene trees, their branch lengths, sequence evolution parameters (including rates of evolution), and population size. 
Due to the large parameter space, co-estimation methods have remained unable to scale to large or even moderate-size datasets despite recent progress \citep{Ogilvie2017} and the use of divide-and-conquer \citep{Zimmermann2014}.

Scalable alternatives to co-estimation are of two types:  summary methods and site-based methods. 
Site-based methods \cite[e.g.,][]{SVDquartets,snapp,pomo} go directly from gene data to the species tree, without inferring gene trees, yet accounting for MSC. 
For example, SVDQuartets, a leading site-based method, uses invariants on site pattern matrices. 
Due to their reduced number of parameters, site-based methods are more scalable than co-estimation \citep{Molloy2019}.

Summary methods divide the inference into two steps (Fig.~\ref{fig:methods}); first, infer gene trees independently for all loci, then, combine these gene trees to get the species tree. 
Under the MSC model, sequence data from different genes are independent {\em conditioned} on gene trees but are not independent for unknown gene trees. 
Thus, summary methods can be understood as ignoring the dependence between gene loci in the gene tree inference step. 
Once gene trees are inferred, combining them to infer a species tree needs specific methods that are statistically consistent under the MSC model. 
Examples of such consistent methods include  STAR \citep{star}, BUCKy-population \citep{Larget2010}, GLASS \citep{glass},
MP-EST \citep{mpest}, STELLS \citep{stells}, 
DISTIQUE \citep{distique}, NJst \citep{njst}, and a related method ASTRID \citep{astrid}.
By breaking the analysis into many independent inferences, the summary approach can produce a very scalable pipeline (requires careful choices of methods). 
Perhaps because of their scalability, summary methods are widely used in  biological analyses~\citep{Molloy2019}. 
In particular, the summary method ASTRAL \citep{astral} has been used in many publications.
In this Chapter, I focus on ASTRAL, intending to give guidelines to practitioners.

Section~\ref{sec:alg}  overviews algorithmic details and theoretical properties of ASTRAL.
Sections~\ref{sec:results} and~\ref{sec:runningtime} summarize the literature on the accuracy and scalability of ASTRAL.
Since the accuracy of ASTRAL depends on its input, Section~\ref{sec:input} is dedicated to best practices in preparing  the input gene trees.
Sections~\ref{sec:output} and \ref{sec:followup} elaborate on the output of ASTRAL and follow-up analyses that can help researchers better understand the results.

\section{ASTRAL Algorithm}\label{sec:alg}
\subsection{Motivation and History}
Computing the probability of a gene tree given a species tree is computationally challenging \citep{Degnan2005}, especially when the gene tree does not have branch lengths in coalescent units.
Thus,  developers of summary methods have looked beyond likelihood-based approaches.
A helpful feature of MSC is that for rooted gene trees with three species (triplets) or unrooted gene trees with four species (quartets), the species tree topology is the most probable gene tree topology \citep{Pamilo1988,Allman2003}.
Thus, on triplets/quartets of species, we can  count the number of rooted/unrooted gene trees and pick the most frequent one as the species tree; it is trivial to show this method is statistically consistent assuming gene trees are sampled from the distribution defined by MSC on a species tree. 
In contrast to triplets and quartets, in the general case of  more species, the species trees can be discordant with the most likely gene trees \citep{Degnan2006,Degnan2009}, a condition known as the anomaly zone. 
 
Several methods have  extended the most-frequent-gene-tree method to more species by decomposing a dataset of $n$ species to all possible ${n}\choose{3}$ triplets or  ${n}\choose{4}$ quartets.
 \cite{Larget2010} suggested using Bayesian concordance factors \citep{Ane2007} to compute the most frequent quartet tree for all possible choices of quartets, and then, combining the quartets using a quartet-joining method \citep{Ma2008}.
More recently, \cite{distique} derived a consistent distance estimate between pairs of species based on how many times they are sisters among all possible quartets that include the two species of interest. 
Instead of finding the highest frequency gene tree, \cite{mpest} defined the pseudo-likelihood of the species tree by decomposing it into all possible triplets, computing the likelihood for each triplet, and combining the likelihoods by assuming independence.
ASTRAL, too, decomposes gene trees to quartets. 

The main insight behind ASTRAL is to realize that the solution to the following optimization problem is a consistent estimator of the species tree (easy to prove based on results of \cite{Allman}).  
Let $\qs{\TR}$ be the set of all quartet tree topologies induced by a tree $\T$.

\begin{problem}[Maximum Quartet Support Species Tree (MQSST)]
Given a set  of $k$ unrooted gene tree topologies $\GT$ on (subsets of) $n$ species, find the species tree $\ST$ that shares the maximum total number of quartet trees with the set of gene trees.
That is, find $\ST = \argmax_{\TR} S(\TR)$ where
\begin{equation}
    S(\TR) =  \sum_{G\in\GT} |\qs{\TR} \cap \qs{G}| \; .
\end{equation}
\end{problem}

MQSST has been studied even before its connection to MSC was realized. 
The problem is NP-hard in several variations \citep{Steel1992b,jiang_ptas,Lafond2016}, but heuristic solutions exist \citep[e.g.,][]{Avni2015}.
One way to achieve scalability is to define a constrained version of the problem.
\begin{problem}[Constrained MQSST]
Solve the MQSST problem such that every branch (i.e., bipartition) of the species tree is drawn from a given set $\X$ of possible branches. 
\end{problem}
\cite{Bryant2001} were the first to define this problem (to my knowledge), which they solved using dynamic programming in time that grows as $O(n^5k+n^4|\X|+|\X|^2)$.
ASTRAL uses a dynamic programming algorithm similar (but not identical) to that of \cite{Bryant2001}, with an improved running time (we were unaware of the method by \cite{Bryant2001} in our original publication.) 
Crucially, solutions to constrained MQSST are consistent estimators under MSC model \citep{astral}.

\subsection{ASTRAL Algorithm}\label{sec:astral}
ASTRAL has three published versions: ASTRAL, ASTRAL-II, and ASTRAL-III, and most recently, a parallel implementation, ASTRAL-MP.
Below, when not otherwise specified, we discuss ASTRAL-III. Readers not interested in mathematical and algorithmic details can skip this section.

\subsubsection{Weight Calculation and Dynamic Programming}
\begin{figure}[t!]
\centering
\includegraphics[width=1\textwidth]{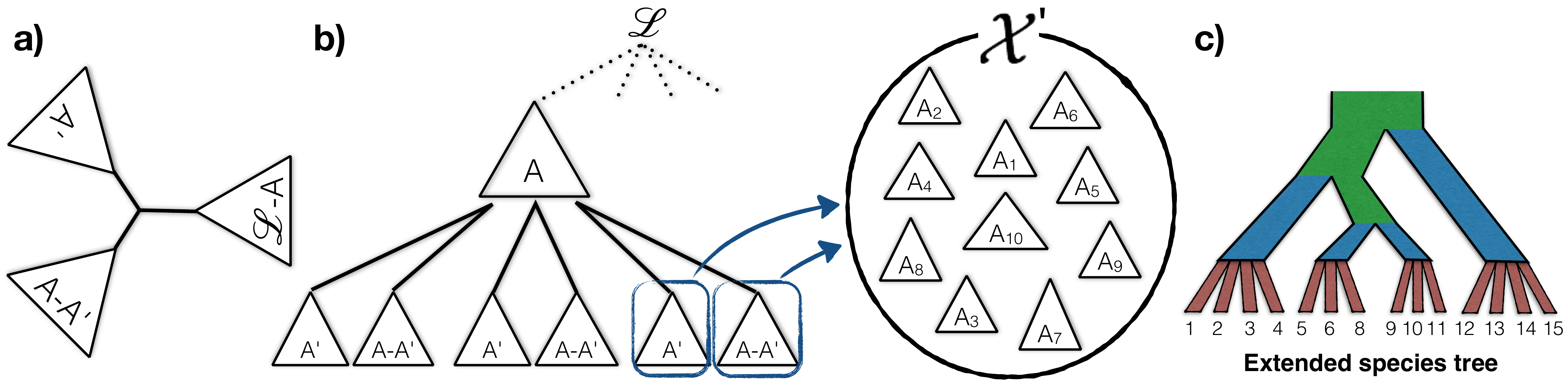}
\caption{(a) An internal node in an unrooted tree creates a (tri)partition of leaves. (b) The dynamic programming recursively divides each cluster $A$ into smaller cluster, drawing possible subsets from the set $\X$. 
(c) An extended species tree; species terminal branches are in blue; individuals are added as polytomies. }
\label{fig:tri}
\end{figure}

A node
in a binary (or multifurcating) unrooted tree $\TR$ corresponds to a partition of leaves into three (or more) parts (Fig.~\ref{fig:tri}a).
Thus, a binary (multifurcating) tree can be represented as a set of tripartitions (multipartitions), one per node. 
The ASTRAL algorithm is based on three insights:
\begin{enumerate}
    \item The number of quartet trees shared between two trees equals half the sum of the number of quartet topologies shared among all pairs of tripartitions/multipartitions, one from each tree.
    \item The number of quartet topologies shared between a tripartition and a multipartition can be computed efficiently {\em without} listing all $n \choose 4$ quartet topologies.
    \item Dynamic programming can be used to find a set of tripartitions that can be combined into a fully binary tree and in total have the maximum possible number of shared quartets with gene trees.
\end{enumerate}

Let the set of species be $\lset$.
Let also $\mathcal{N}(\TR)$ be the set of internal nodes in a tree $\TR$, represented as  multipartitions. 
Any (species) tree that includes a tripartition $P$ as a node shares a certain number of  quartet topologies with any (gene) tree that includes a multipartition $M$ as a node; we let $QI(P,M)$ denote this quantity.
We define the {\em weight} of a tripartition as:
 \begin{equation}
 w(P)=\frac{1}{2}\sum_{G\in\gset}\sum_{M\in \mathcal{N}(G)}QI(P,M) \; .\label{eq:w}
 \end{equation}
Insight 1 asserts that  
$
S(\TR)=\sum_{P\in \mathcal{N}(P)}  w(P)
$.
Thus, we need to $i)$ compute $w(P)$ efficiently, and $ii)$ find the tree with the maximum sum of $w(P)$ values. 
\cite{astral3} derived an efficient formula for $QI$. 
Given a multipartition $M=M_1| \ldots | M_d$ (representing an internal node in a gene tree)
and a tripartition $P=P_1| P_2| P_3$ (an internal node in a species tree), for  $1 \leq i \leq d$ and  $1\leq j,k \leq 3$,  let $\inters{j}{i} =|M_i \cap P_j|$, $S(j)=\sum_{i=1}^d \inters{j}{i}$, and $R({j,k}) =\sum_{i=1}^d \inters{j}{i} \inters{k}{i}$.
Let $\bigl( \begin{smallmatrix}2 & 1 & 1\\ 3 & 3 & 2\end{smallmatrix}\bigr)=(\oth{i}{j})$ be a constant matrix. Then, 
\begin{small}
\begin{equation}\label{eq:npol}
QI(P,M)=
\frac{1}{2} \sum_{i=1}^d\sum_{j=1}^3  \binom{\inters{j}{i}}{2} \Big(\big(S({\oth{1}{j}}) - \inters{\oth{1}{j}}{i}\big) \big(S({\oth{2}{j}}) - \inters{\oth{2}{j}}{i}\big) - R({\oth{1}{j},\oth{2}{j}}) +  \inters{\oth{1}{j}}{i} \inters{\oth{2}{j}}{i}\Big) \; .
\end{equation}
\end{small}

Computing this equation requires $\Theta(d)$ time given $\inters{i}{j}$ values.
ASTRAL-III uses a polytree data structure to represent gene trees such that each $\inters{i}{j}$ can be computed in constant time. 
The data structure also compresses nodes that appear in multiple gene trees. 
\cite{astral3} showed how to compute $w(P)$ in $\Theta(D)$ where $D=O(nk)$ is the sum of the cardinalities of unique partitions observed in all gene trees.

To maximize $S(T)$,  ASTRAL recursively divides $\lset$ into two smaller subsets (called {\em cluster}s) until it reaches leaves (Fig.~\ref{fig:tri}b). 
Each cluster is divided such that  the total sum of the weights below it is maximized among allowable divisions. 
This recursive method solves the MQSST problem optimally if all ways of dividing a cluster $A$ into $A'\subset A$ and $A\setminus A'$ are examined. 
But that approach has an exponential running time, hence, the need for the constrained MQSST problem.
Assume we have defined a set $\X$ of allowable bipartitions for the species tree. 
Let  
$\X'=\{A: A|\lset \setminus A\in \X\}$ and $Y=\{(C,D): C\in \X', D\in \X', C\cap D = \emptyset, C \cup D \in \X'\}$. 
\cite{Kane2017} showed $|Y|< |\X'|^{1.726}$.
We restrict the dynamic programming such that $(A',A \setminus A')\in Y$ (Fig.~\ref{fig:tri}b).
Let $S^*(A)$ be the score 
for an optimal subtree on cluster $A$.
Then, the following dynamic programming solves the constrained MQSST problem optimally in time that scales in the worst case as $O(D|\X|^{1.726})$, spending the majority of time in computing Eq.~\ref{eq:w} 
\citep{astral3}:
\begin{equation}
\label{eq:dp}
S^*(A) = 
\max_{(A',A \setminus A') \in Y} S^*(A') + S^*(A \setminus A') + w(A'|A\setminus A'|\lset \setminus A) \; .
\end{equation}

\subsubsection{Constraint set}
The sufficient condition for ASTRAL to be statistically consistent under the MSC model is to have all bipartitions from input gene trees in the set $\X$ \citep{astral}. 
However, \cite{astral2} 
showed that $\X$ might need to be expanded in order to obtain high accuracy in practice.
ASTRAL-III expands $\X$ using rules summarized below, but users can also directly expanded the set.
These heuristics rely on a similarity matrix computed based on how often a pair of species appear as sisters in gene tree quartets. 
\begin{itemize}
    \item When input gene trees are incomplete, first complete them before adding their bipartitions to $\X$ using the similarity matrix \citep{astral2}. Similarly, when gene trees include polytomies, first resolve their polytomies in several ways before adding the bipartitions to $\X$ \citep{astral3}. 
    \item Compute greedy consensus trees of gene trees with various thresholds (the minimum required frequency for adding bipartitions to the consensus), resolve the polytomies in the greedy consensus trees, and add the resulting bipartitions to $\X$. 
    To resolve polytomies, subsample one species from each side of the polytomy, and resolve it using two approaches: using the similarity matrix and by computing greedy consensus trees on the subsampled taxa. 
    \item Ensure  heuristics do not add more than $O(nk)$ bipartitions. 
    With this rule, we get running times that increase as  $O(D(nk)^{1.726})=O((nk)^{2.726})$.
\end{itemize}

\subsubsection{Multiple individuals}

ASTRAL can easily be extended to inputs where more than one species represent each species. 
 \cite{Allman} have introduced the concept of an extended species tree: start with the species tree, and for each species, add all individuals sampled from that species under it, creating polytomies when needed (Fig.~\ref{fig:tri}c). 
\cite{astral-multi} have extended dynamic programming of Equation~\ref{eq:dp} to compute the optimal extended species tree given gene trees with multiple individuals from some or all species. 
The dynamic programming is unchanged (treating individuals as taxon set $\LS$), except for two modifications.
    $(i$) The boundary conditions need to change such that the algorithm stops as soon as a cluster equals the set of individuals of a species.
    $(ii)$ Set $\X$ needs to change such that each cluster has either all or none of the individuals of each species. Satisfying this condition required new methods for building  $\X$ \citep{astral-multi}. 


\subsection{Summary of known theoretical results related to ASTRAL }

\begin{description}
\item[Consistency - general.]
All versions of ASTRAL give a statistically consistent estimator of the species tree if input gene trees are sampled randomly under the multi-species coalescent model (i.e., with no gene tree error, no sampling bias, and no model violations). 
\item[Consistency - missing data.]
{\cite{Nute2018} showed that ASTRAL remains statistically consistent when species are allowed to be missing from gene trees. 
For the exact version, key required assumptions are that the presence of a gene for a species should be independent of the gene tree topology and presence of other genes for that species. 
The default (constrained) version is also consistent if each clade of the species tree has a {\em non-zero} chance of having no missing data in each gene. }
\item[Inconsistency - estimated gene trees.]
{
\cite{Roch2018} have proved that  ASTRAL and other ``reasonable'' summary methods that use gene tree topology are statistically inconsistent if each gene has limited length and gene trees are computed using ML. 
Under specific conditions, they show ASTRAL and even partitioned ML fail due to long branch attraction even if there is {\em no gene tree incongruence}.} 
\item[Inconsistency - Reticulation.] \cite{Solis-Lemus2016} have shown that ASTRAL can be statistically inconsistent under certain conditions when gene trees evolve on a phylogenetic network (thus, with a combination of ILS and gene flow). 
\item[Sample complexity.] 
\cite{Shekhar2017} have shown that the number of genes required by the exact version of ASTRAL to compute the correct species tree with high probability grows quadratically with the inverse of the shortest branch length and grows logarithmically with the number of species. 
\end{description}

\section{Accuracy}\label{sec:results}



\begin{figure}[p]
\centering
\includegraphics[width=1\textwidth]{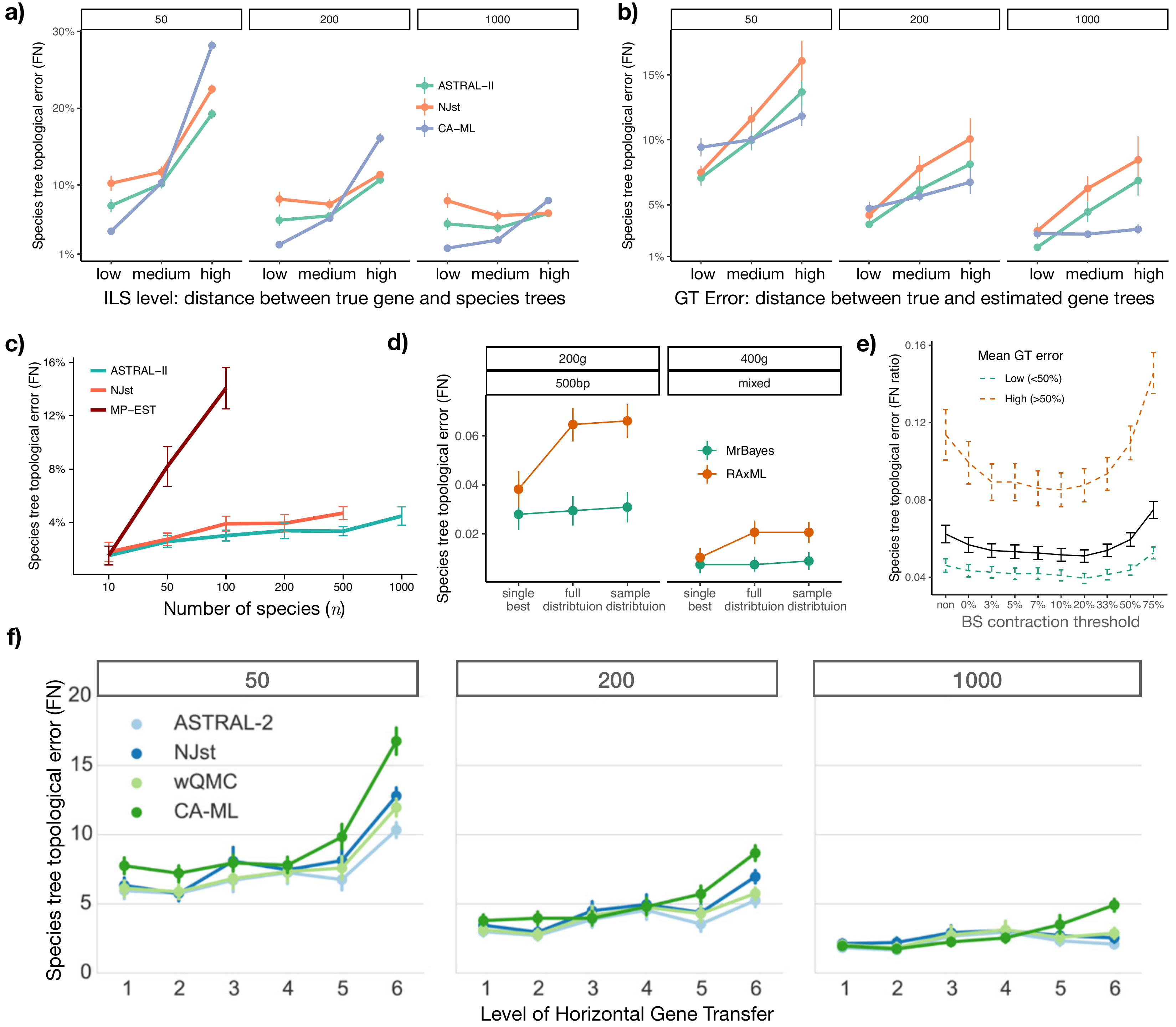}\vspace{-5pt}
\caption{{\bf Accuracy of ASTRAL in simulations}.
Shown is the species tree error, defined as the proportion of branches in the true tree missing from the estimated trees; i.e., False Negative (FN) rate. 
(a-c) Simphy simulations by \cite{astral2} comparing concatenation (CA-ML) and  summary methods.
(a) The level of ILS is set to low, medium, or high by adjusting tree height ($10^7$, $2\times 10^6$, $5\times 10^5$ generations) and affects mean RF distance between true species trees and true gene trees (9\%, 34\%, 68\%).
Speciation rate: $10^{-7}$; number of genes: 50, 200, or 1000 (boxes); 201  species.
(b) Replicates in (a) are categorized into three sets based on the mean normalized RF distance between true gene trees and gene trees estimated by FastTree: $[0,25]\%$, $(25,40]\%$, $(40,100]\%$, corresponding to low, medium, and high gene tree error.
(c) Error versus the numbers of species for medium levels of ILS ($2\times 10^6$ height) and speciation rate $10^{-6}$ with 1000 genes.
(d) Mammalian-like simulations from \cite{astral} with 37 species with 200 gene trees (500bp), or 400 gene trees (mix of 500bp and 1000bp).
Gene trees estimated using RAxML or MrBayes. 
Input:  {\em single best} tree per gene (ML for RAxML or maximum credibility for MrBayes); 
{\em full distribution} per gene (200 BS replicates for RAxML or a sample of 200 trees for MrBayes); 
{\em sample distributions} to get a single tree per gene and repeat 200 times to report their consensus (a.k.a MLBS).
MrBayes results are unpublished.  
(e) 
Simulations with 100 species and moderately high ILS (46\% RF between true gene trees and species trees) by \cite{astral3}.
Contracting branches with $\leq20$\% BS support (threshold of contraction shown on x-axis)  from best ML trees reduces error.
Dividing replicates into high and low gene tree error shows that contraction helps mostly in the high error case.
(f) 
50 species simulations by \cite{astral-hgt}, comparing accuracy of methods in presence of HGT.
All six model conditions (x-axis) have ILS but differ in level of HGT, ranging from no HGT (1) to very high (6). 
Thus, true gene tree discordance varies: $\sim$33\% (1-3), $\sim$45\% (4), $\sim$55\% (5), and $\sim$70\% (6).
CA-ML is the least robust and ASTRAL is the most robust to HGT. 
}
\label{fig:astral}
\end{figure}

\paragraph{Versus concatenation.}
{Simulation studies have indicated that the relative performance of concatenation and ASTRAL depends at least on two factors: the amount of true gene tree discordance (ILS) and the amount of gene tree estimation error \citep[e.g.,][]{astral,astral2,Giarla2015,Molloy2018}. }
For example, \cite{astral2} found ASTRAL to be more accurate than concatenation using ML (CA-ML) when true gene tree discordance (i.e., level of ILS) was high  (Fig.~\ref{fig:astral}a) or when gene tree error was relatively low  (Fig.~\ref{fig:astral}b).
In contrast, CA-ML was more accurate when either the true gene tree discordance was low or when the discordance was moderate or high, but the gene tree error was also high. 
The two methods had similar accuracy when ILS levels were moderate, {\em and} gene tree error was also moderate; e.g., when normalized \cite{rf} (RF) distance between true and estimate gene trees was between 20\% to 40\% (Fig.~\ref{fig:astral}b).
Moreover, \cite{astral-hgt} found ASTRAL outperforms CA-ML in the presence of both ILS and HGT (Fig.~\ref{fig:astral}f).

{In practical terms, when gene tree discordance is very low or when gene tree error is expected to be high, CA-ML may be preferable to ASTRAL whereas in other scenarios ASTRAL is preferable. }
Because neither method universally dominates the other, it seems wise to use both methods and compare the results. 
One way to decide is to simulate data emulating real data and comparing methods.
For example, \cite{Giarla2015} found that for their dataset of tree shrews, both ASTRAL and CA-ML produced one wrong branch in simulations that emulated the real data, but only CA-ML had high bootstrap support for the wrong branch. 
\cite{Ballesteros2019} showed in simulations that ASTRAL could recover the correct branch even when a branch does not appear in \textit{any} of the input gene trees.

\paragraph{Versus other summary methods.}
Several simulations studies compare summary methods \cite[e.g.,][]{astral,astral2,distique,Molloy2018,astrid}, {including some that do not involve developers of ASTRAL~{\citep{Giarla2015,Ballesteros2019}}}.
{
Overall, the accuracy of ASTRAL has compared favorably to alternative ILS-based summary methods such as NJst \citep{njst}, ASTRID \citep{astrid}, MP-EST \citep{mpest}, wQMC \citep{Avni2015}, as well as consensus and supertree methods such as greedy consensus, MulRF \citep{Chaudhary2013}, and MRP \citep{Ragan1992}. 
For example, ASTRAL outperformed NJst by small but consistent margins in simulations by \cite{astral2} (Fig.~\ref{fig:astral}a-c) and in simulations with HGT done by \cite{astral-hgt} (Fig.~\ref{fig:astral}f) and was essentially tied with ASTRID in \cite{Molloy2018} and \cite{astrid}.
DISTIQUE was close to ASTRAL but not any better in \cite{distique}. 
ASTRAL dominated MP-EST \citep{astral,astral2}, especially with large numbers of species (Fig.~\ref{fig:astral}c).}
\cite{Giarla2015} showed ASTRAL outperforms MulRF in terms of topological accuracy on simulations that match their real dataset of tree shrews. 
Beyond accuracy on conditions that seek to emulate real data,
\cite{Shekhar2017} have compared ASTRAL with NJst in idealized cases in terms of data requirements: the number of error-free genes required to recover the correct species tree with high accuracy.
Their simulations showed mixed patterns: out of three true species trees tested, they found ASTRAL required fewer genes in two cases (with only short branches), and NJst required fewer genes in the third case (with long basal branches). 

\paragraph{Other results.}
Simulation studies have tested the robustness of ASTRAL to factors such as HGT (Fig.~\ref{fig:astral}f), gene flow \citep{Solis-Lemus2016}, gene tree error \citep{WSB}, and missing data \citep{Molloy2018} (discussed later).
\cite{astral-multi} studied the relative impact of increasing the number of loci or the number of individuals per species on ASTRAL accuracy and found more loci to be far more beneficial.
Beyond simulations, researchers have also compared ASTRAL to other methods on empirical data \citep[e.g.,][]{Giarla2015,Simmons2016,Edwards2016,Meiklejohn2016,Streicher2015,Shen2017}.
Since the ground truth is not known on real data, these results are harder to interpret and cannot be easily summarized without loss of important nuance. 
Referring the reader to these publications, I note that 
overall, the performance of ASTRAL on real data has been positive.

\section{Running time}\label{sec:runningtime}

The running time of ASTRAL has improved through its three versions, both in theoretical guarantees of worst-case asymptotic running time and in empirical measures. 
ASTRAL-I had guaranteed polynomial running time but is the slowest version. 
ASTRAL-II did not guarantee polynomial running time but was faster than ASTRAL-I. 
The current version, ASTRAL-III, has an asymptotic worst-case running time of $O(D(nk)^{1.726})$, which itself is $O((nk)^{2.726})$ (recall that $D$ is the sum of degrees of \textit{unique} nodes in input gene trees). 
The theoretical running time, thus, is a function of $n$, $k$, and the amount of gene tree discordance, which controls both the search space and $D$.
In practice, on datasets tested by \cite{astral3}, the empirical running time of ASTRAL-III seems to increase with $n^2k^2$.
Thus, for example, a researcher planning to double the number of gene trees can expect a four-fold increase in the running time. 
Similarly, other things equal, increasing the number of species should roughly quadruple the running time. 

Example numbers may be instructive. 
\cite{astral3} report that ASTRAL-III took roughly 16 hours on average for a dataset with very high ILS, 48 species, and 16,000 genes (so, a large-$k$ scenario), and roughly 9 hours  on a dataset with moderate ILS, 5000 species, and 1000 genes (so, a large-$n$ dataset).
For a comparison of the running time of ASTRAL to other  methods, see \cite{Molloy2019}.\FILL{Chapter \ref{molly-warnow}}

\paragraph{ASTRAL-MP.}
 \cite{astralmp} have recently developed a parallel version of ASTRAL for CPU (multi-core and vectorization) and  GPU that can analyze very large datasets. 
This version, called ASTRAL-MP, speeds up runs by up to 150X compared to ASTRAL-III, especially for datasets with large numbers of gene trees.
On a dataset with 10,000 species, 1000 gene trees, and moderate ILS, ASTRAL-MP takes between 5 to 32 hours (11 hours on average) given a single GPU and 24 cores with AVX2. 
On a real insect transcriptomic dataset 
with 144 taxa and 1478 genes, each with 100 bootstrapped gene trees ($k = 147800$ in total),  ASTRAL-MP with four GPUs and 24 cores finished in 35 hours.

\section{Input to ASTRAL: Practical Considerations}\label{sec:input}
The ideal input to coalescent-based methods is a set of perfectly aligned orthologous regions present in all genomes, with each region small enough to avoid recombination but large enough to have a strong phylogenetic signal and with regions distributed randomly across the genome and placed far enough apart to make them fully unlinked.
For summary methods, gene trees are ideally estimated under models of sequence evolution that are correct (but not overly parameterized) using consistent methods that utilize the data efficiently (i.e., have optimal sample complexity).
Satisfying all these requirements is hard, if not impossible. Thus, phylogenomic projects seeking to use summary methods like ASTRAL face  many practical choices.

\begin{figure}[t!]
\centering
\includegraphics[width=0.95\textwidth]{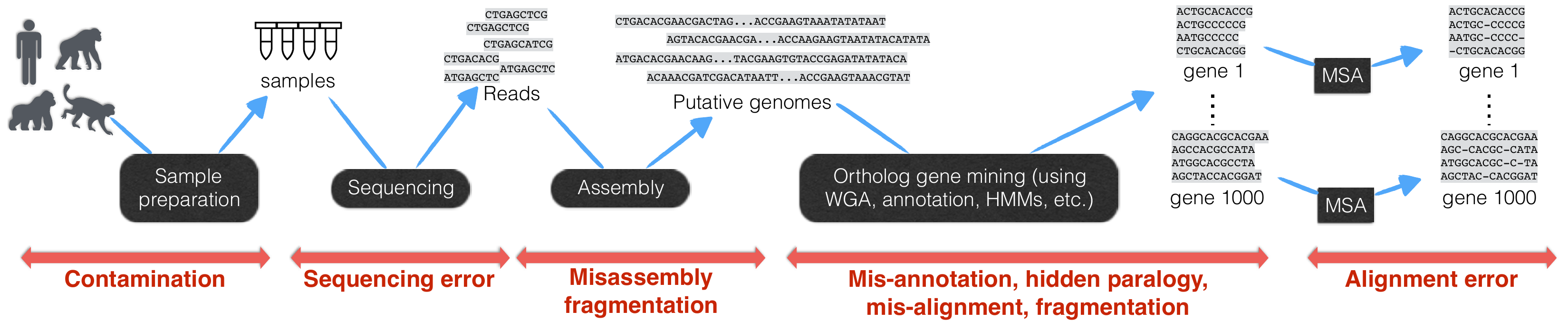}
\vspace{-10pt}
\caption{The phylogenomics pipeline. To get the typical input to coalescent-based methods, a set of gene alignments, many steps have to be taken, and each step is prone to errors that can propagate. }
\label{fig:pipeline}
\end{figure}

In practice, phylogenomic analyses include many steps before arriving at gene tree and species tree estimation (Fig.~\ref{fig:pipeline}). 
From sample preparation to sequencing, assembly, and annotation, to orthology detection and multiple sequence alignment (MSA), the pipeline includes steps that are far from trivial \citep{pitfalls}. 
Each step is error-prone, and some steps (e.g., orthology detection and MSA) seek to solve computational problems that are incidentally best solved with the knowledge of the phylogeny. 

Several groups have discussed error propagation through steps of the pipeline and the impact on the species tree \citep[e.g.,][]{Patel2013,binning,Gatesy2014,pitfalls,Molloy2018}.
We can aspire to move away from a pipeline approach and towards a unified statistical inference, with full join modeling of uncertainty \citep{Szollsi2014}. 
Since this end-to-end co-estimation remains unavailable currently and likely impractical in the near future, we are left having to deal with pipelines, which requires awareness of errors and making an effort to mitigate their impact. 
Below, I discuss best practices that have emerged from published work in preparing the input to ASTRAL.

\subsection{Gene tree estimation}

\subsubsection{Gene tree uncertainty}
The standard input to ASTRAL is ML gene trees inferred under standard models of sequence evolution.
Restricting ourselves to ML, several options are available.

\begin{description}
\item[bestML:] The most straightforward choice is to use the tree with the best likelihood found by a heuristic ML method. 
The bestML input is the most natural approach but ignores gene tree uncertainty. 
\item[Contracted bestML:] Each gene tree is bootstrapped, and support values are computed for bestML trees. 
Then, branches with extremely low support in bestML trees are contracted, and the resulting multifurcating trees are used as input to ASTRAL (one per gene).  
\item[MLBS:] Multi-locus Bootstrapping (MLBS) seeks to model uncertainty by performing bootstrapping for each gene. 
Then, these bootstrapped gene trees are used to create several inputs to ASTRAL (with or without gene resampling); running ASTRAL on each input set produces a set of outputs, which are then summarized using methods such as greedy consensus to generate a final consensus result. 
\item[ALLBS:] All replicate bootstrapped gene trees are combined to form a single input to ASTRAL.
\end{description}

{At least two simulation studies \citep{mrl-sysbio,astral} have shown that MLBS or ALLBS have lower accuracy than simply using bestML, except perhaps when only a small number of genes are available (see Fig.~\ref{fig:astral}d). 
\cite{localpp-sm} have provided an explanation. 
The set of bootstrapped gene trees show a higher level of gene tree discordance than the set of bestML gene trees.
The increased discordance is not biological but is a result of the lowered phylogenetic signal in bootstrapped gene alignments. 
This increased level of gene tree error, as we previously saw (Fig.~\ref{fig:astral}b), can reduce the accuracy of ASTRAL.}

Contracted bestML, in contrast to MLBS, can improve accuracy. 
The important (if somewhat counter-intuitive) point to remember is that only collapsing branches with {\em extremely low} support improves accuracy, and contracting other branches can \textit{increase} error. 
\cite{astral3} have shown that collapsing branches with BS below 5-20\% can improve accuracy by a substantial margin (Fig.~\ref{fig:astral}d) and that the improvements in accuracy are higher when input gene trees have higher levels of error and when more gene trees are available.
For example, for input gene trees with the mean error above 50\%, 
ASTRAL tree enjoys a 25\% reduction in error (from 0.114 to 0.085) after contracting gene tree branches with support below 10\%. 
Thus, contracting {\em very} low support branches can increase accuracy substantially. 
Aggressively collapsing branches with support $<50$\% or $<75$\% (i.e., only keeping high support branches) can substantially reduce the accuracy (Fig.~\ref{fig:astral}d). 
Future research should explore smarter algorithms for collapsing low support branches. 

\cite{GGI} have suggested inferring gene trees constrained to include a set of predefined undisputed clades chosen by the researcher. 
They show promising results on several datasets using ASTRAL applied with such constrained gene trees. 
However, the method can also remove some of the real discordances among gene trees (as opposed to noise). 
As pointed out elsewhere \citep{Mirarab2017}, this approach runs several theoretical risks, including biasing result in unexpected ways. 

\subsubsection{Inference tools and models}
{
The choice of the gene tree inference tool may be consequential, and
published simulations have compared FastTree \citep{fasttree-2} and RAxML \citep{Stamatakis2014}. 
Despite earlier results \citep{fasttree-raxml},
\cite{insects} have found using simulated and empirical data that FastTree \textit{can} be less accurate than RAxML in inferring gene trees (under limited conditions they test), and the increased gene tree error leads to less accurate ASTRAL trees. 
Thus, using best available ML methods (ideally with multiple starting trees) is preferable to faster tools in biological analyses.} 
Moreover, effects of misspecified sequence evolution models have been discussed for phylogenomics in general \citep[e.g.,][]{Phillips2004,Jeffroy2006}, but to my knowledge, have not been studied for ASTRAL.
We should expect that systematic model misspecification can lead to biases in estimated gene tree distributions, which can lead to errors in the species tree. 

In unpublished simulations, I have compared the ML method RAxML and the Bayesian method MrBayes \citep{mrbayes} on a mammalian-like simulated dataset \citep{astral}. 
Like ML, the output of MrBayes is used in three ways: using a single maximum credibility tree per gene, using a large sample of trees per gene (akin to ALLBS), and repeatedly sampling single trees from gene tree distributions produced by MrBayes (akin to MLBS).
Interestingly, unlike ML and BS, the use of Bayesian distributions removes the sensitivity to the mode of input so that all three types of input perform similarly (Fig.~\ref{fig:astral}d). 
These results also show a small advantage in using MrBayes compared to RAxML when both are using in their best setting (i.e., a single tree per gene). 
These preliminary results warrant more studies in the future.

\subsection{Filtering data}

A vexing problem in phylogenomics is that curating sequence data using visual inspection is impossible. 
Thus, methods for \textit{detecting} errors automatically, perhaps in downstream steps, are also needed.
Empirical studies often employ several mostly {\em ad hoc} methods for filtering erroneous data \citep{pitfalls}, typically relying on a mix of visual inspection of (parts of) data and automatic error detection tools. 
While the extent of the negative impact of errors in input on the output ASTRAL tree is not fully understood, efforts to minimize such errors seem necessary. 
However, tampering with data to remove error can also remove signal and introduce bias -- and thus warrants caution and  careful study.

\subsubsection{Filtering leaves from gene trees}

One way of detecting abnormalities is to examine the estimated gene trees that show unexpected patterns.
For example, \cite{1kp-pilot} rooted gene trees and detected and removed branches from gene trees with extremely long root to tip distance compared to the other species. 
Visual inspection of gene trees is also what Gatesy, Springer, and colleagues have used in several of their published criticism of previous phylogenomic studies \citep{Gatesy2014,SpringerGatesy,Springer2015,Springer2017}.  
A well-studied automated approach for detecting species with unstable positions in individual gene trees is rogue taxon detection ~\citep[e.g.,][]{Aberer2013,Westover2013}.
Rogue taxon detection methods tend to identify the same species (usually those on  long branches) on many genes \citep{TreeShrink}.
Since removing the same taxon from many genes reduces taxon occupancy, rogue taxon removal may prove problematic. 
The effect of rogue taxon removal on ASTRAL, to my  knowledge, has not been tested.

\begin{figure}[!tb]
\centering
\begin{footnotesize}
    \includegraphics[width=0.85\textwidth]{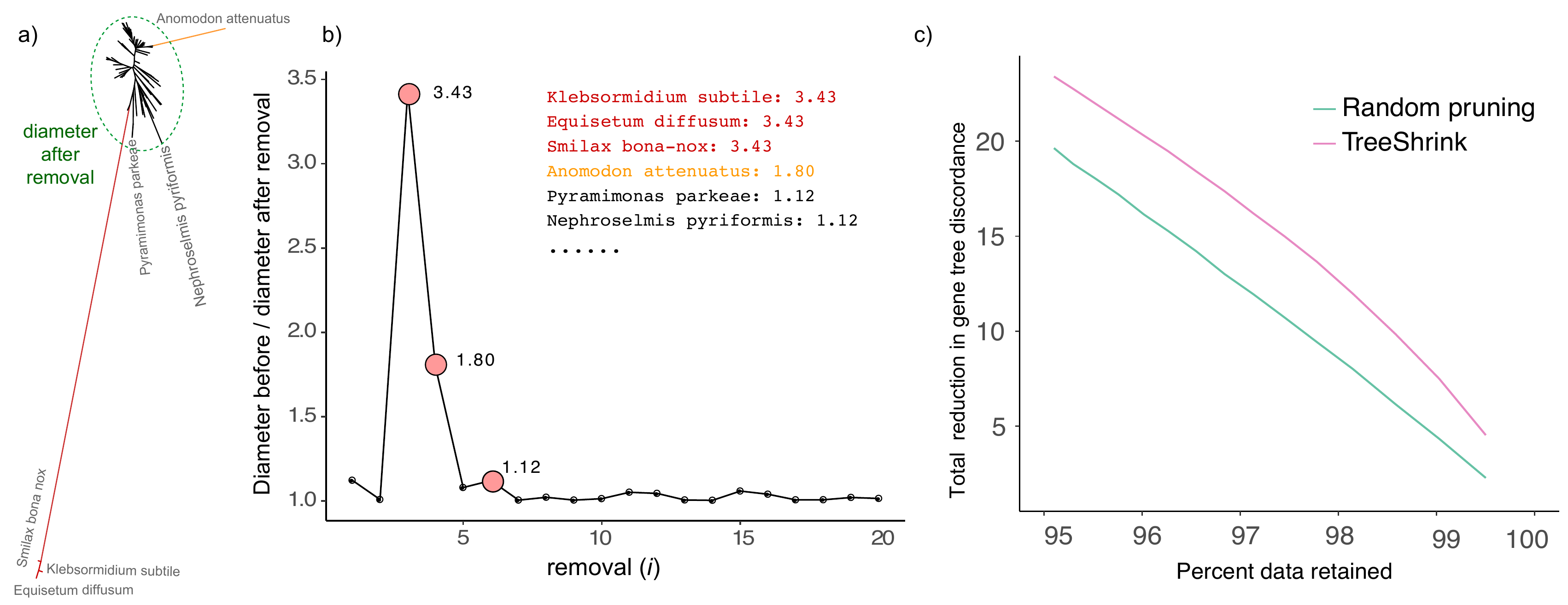}
 \end{footnotesize}
  \caption{
  {\bf TreeShrink} on \cite{1kp-pilot} dataset. 
  (a) A gene tree with abnormally long branches.
(b) TreeShrink identifies three (red) tips that increase the diameter 3.4X and are likely erroneous, followed by another (orange) that {may} also be problematic. 
(c) TreeShrink reduces discordance substantially more than random filtering.
y-axis: \textit{reduction} in gene tree discordance defined as mean pairwise matching splits distance between gene trees. 
}\label{fig:treeshrink}
\end{figure}

\paragraph{TreeShrink.}
\cite{TreeShrink} developed an automatic method called TreeShrink to find suspicious patterns of branch length in gene trees (Fig.~\ref{fig:treeshrink}a). 
TreeShrink tries to successively shrink the diameter (i.e., the maximum total branch length between any two leaves)  of each gene tree by removing species. 
Then, for each species in each gene tree, TreeShrink computes a \textit{signature} (Fig.~\ref{fig:treeshrink}b), which quantifies its impact on the diameter of the gene tree. 
Finally, it examines the \textit{distribution} of signatures of each species across all gene trees and detects outliers in these distributions using a simple heuristic. 
Outliers is a case when a species has an uncharacteristically large signature (e.g., is on a long branch) in a gene tree compared to the rest of gene trees.
Since outliers are defined using a distribution across genes for a single species, a taxon with high signatures in most genes will not be detected as an outlier in those genes (outgroups tend to be like this). 
In contrast, a taxon with low signatures in most genes but high signatures in a handful of genes will be detected as an outlier for those high-signature genes. 
Once the abnormally long branches are detected, TreeShrink removes specific species from specific genes but does not remove the entire gene. 
\cite{TreeShrink} showed on several biological datasets that TreeShrink reduces the pairwise discordance among gene trees beyond random removal of taxa (Fig.~\ref{fig:treeshrink}c) and also often beyond methods such as RogueNaRok \citep{Aberer2013}.
Moreover, it avoids removing any specific species from too many genes.

\subsubsection{Filtering entire gene trees}

A more extreme form of filtering is to remove entire gene trees. 
Several criteria for removing entire genes have been proposed and tested~\cite[e.g.,][]{Meiklejohn2016,Hosner2016,Chen2015,Huang2016,Longo2017,Blom2017}. 
Some of the criteria have to do with missing data and will be discussed later, while  others relate to error and uncertainty in gene trees.  
{
\cite{Molloy2018} give a recent summary of the literature and also, to my knowledge, give the only study that evaluates the accuracy of ASTRAL in simulation in response to removing full genes. 
Their simulations show that removing genes with high gene tree estimation error can improve accuracy for low levels of ILS but {\em reduces} accuracy for moderate to high levels of ILS. 
Nevertheless, even for ILS levels where removing genes helps accuracy, on real data, we do not have direct access to gene tree error to decide what genes to remove. 
Instead, we have to resort to filtering by other factors, such as bootstrap support. 
Since these proxies do not perfectly correlate with gene tree error, the positive impact of filtering may further diminish.
Neither \cite{Molloy2018} nor any other simulation study tested filtering by proxies for ASTRAL. 
\cite{Lanier2015} and \cite{Liu2015b} have studied the question for STEM and MP-EST and have observed little or no reason for filtering.
On empirical datasets, the conclusions have been mixed, with some studies \cite[e.g.,][]{Hosner2016,Meiklejohn2016,Longo2017} recommending removal of gene trees and others finding no evidence that filtering helps~\citep{Chen2015,Blom2017}.
Overall, there is little evidence in the literature suggesting that removing entire genes because of lack of support is helpful to ASTRAL analyses}. 


\subsubsection{Filtering for missing data}
The most common type of filtering is in response to missing data. 
For summary methods,  two types of missing data exist --
\textit{missing genes} and \textit{fragmentary data} (\textit{type I} and \textit{type II} in the parlance of \cite{Hosner2016}). 
The two types have different consequences and have inspired two types of filtering.

\begin{description}
    \item [Type I (missing genes).]
{This type of missing data occurs when a gene is entirely missing for some of the species but is present in others.
    The only suggested filtering for these types of missing data is to remove them. \cite{Molloy2018} found no evidence  in simulations that removing genes with this type of missing data helps accuracy. 
    Their results are in agreement with several empirical studies \cite[e.g.,][]{Chen2015,Hosner2016} that also saw no benefit in filtering genes.} 
    \item [Type II (fragmentation).] When a species includes a gene, but only partially, it introduces missing data in the gene tree inference step. 
    Both  \cite{Hosner2016} and \cite{insects} showed that the presence of fragmentary data can be problematic, confirming earlier observations \cite[e.g.,][]{1kp-pilot,Springer2015}. 
    They show fragmentary sequences  increase gene tree error, which translates to increased species tree error.  
    \cite{insects} suggested a simple yet effective solution: remove  species with fragmentary sequences from gene alignments before inferring the gene tree. 
    The  optimal level of filtering depends on the dataset; \cite{insects} defined species with less than half of the total alignment length as fragmentary while \cite{1kp-pilot} used a one-third threshold. 
\end{description}

To summarize, removing loci because of missing species is not recommended, but removing specific species from loci because of fragmentation  is recommended.
{Note that filtering fragmentary data replaces type II missing data with type I. 
That such a trade-off helps accuracy, once again, underscores the negative impact of gene tree error and the benefit in reducing error -- even if this reduction adds to missing data.}
Note that the distinction between types I and II is not relevant for concatenation.
There is no reason to think that removing fragmentary data from a concatenation analysis could help accuracy, as it only adds missing data.

Results showing that removing type I missing data fails to help accuracy do not imply  missing genes are harmless.
As shown in simulations by \cite{Nute2018}, missing genes can increase error in ASTRAL trees, especially when the number of genes is low, the amount of ILS is very high, or when entire clades tend to be missing. 
Thus, missing data can hurt accuracy, but filtering low occupancy loci is not a solution.

Recently,  \cite{Gatesy2018} added a twist. For a set of empirical data, given two alternative species trees, they computed the difference in quartet score of the two trees \textit{for each gene} and called it partitioned coalescence support (PCS). Genes with extremely high PCS for either alternative tree \textit{tend to} be a lot more complete than other genes. One problematic observation is that in some cases, ASTRAL trees change if only a couple of these high PCS genes are removed. Thus, in the presence of uneven taxon occupancy, results may be driven by a handful of genes, an observation that makes sense given the rapid growth of the number of quartets as trees become larger. These results suggest that perhaps, contrary to common wisdom, having genes with similar levels of occupancy is more important than avoiding missing data. Future work should further explore the implications of these results. 


\section{ASTRAL Output}\label{sec:output}

\subsection{Species tree topology and its quartet score}
ASTRAL outputs the tree with the maximum quartet score among all trees within its search space (defined by $\X$). 
Since ASTRAL limits the search space (unless run with {\tt -x}), it is possible that other trees with better quartet scores exist. 
In simulations, increasing the search space beyond the default used in ASTRAL-II or -III (e.g., by adding all bipartitions from the true tree) rarely even improves accuracy, though it occasionally improves the quartet score slightly \citep{astral2,astral3}.

ASTRAL is a statistically consistent estimator under the MSC model given gene trees sampled from the true distribution defined under MSC.
However, ASTRAL does not use a parametric model and is not tied to likelihood under the MSC model. 
As such, ASTRAL can be considered a non-parametric estimator.
{As it has been long argued \citep{Holmes2003}, absent access to the correct model, reliance on non-parametric methods can be beneficial.}
Thus, ASTRAL (and other non-parametric methods like NJst/ASTRID) may be more robust than parametric methods (e.g., MP-EST), especially given the limited reach of the MSC model.
After all, MSC ignores {\em both} gene tree error \textit{and} biological sources of discordance other than ILS.  
ASTRAL is a natural estimator for any model where the most likely gene tree matches the species tree for quartets, as is the case for some HGT regimes \citep{Roch2013}. 
Further, ASTRAL has even been used as a supertree method outside the phylogenomics context, with reasonable results \citep{fastRFS}.

Along with the tree topology, ASTRAL outputs its quartet score, which  is the number of quartet trees in gene trees that are present in the species tree. 
We normalize the absolute value by the total number of quartet trees in input gene trees (e.g., $k {n \choose 4}$ if there is no missing data) to give a more interpretable score. 
For example, a quartet score of 0.8 means that 80\% of quartet trees in input gene trees are in the output tree. 
Thus, the normalized quartet score can be used as a measure of the amount of gene tree discordance. However, the score has to be interpreted with care as gene tree error  is likely to reduce the quartet score.

\subsection{Branch Lengths in Coalescent Units}
\begin{figure}[tbp!]
\centering
\includegraphics[width=.9\textwidth]{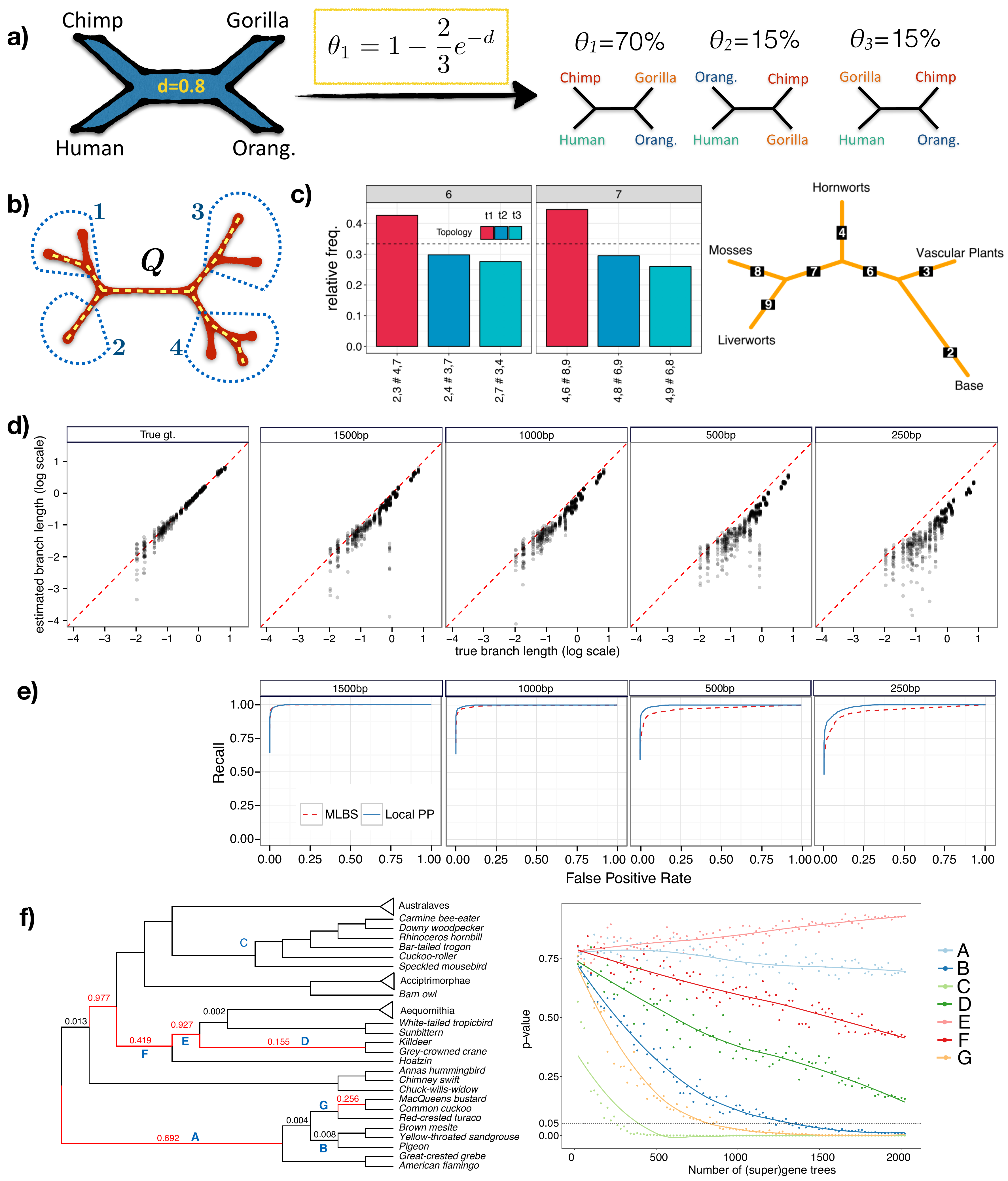}
\vspace{-5pt}
\caption{{\bf Per-branch quartet frequencies}.
(a) The distribution of quartet gene tree topologies ($\theta_{1},\theta_{2},\theta_{3}$) as a function of the CU branch length $d$. 
(b) Branch $Q=12|34$ can be rearranged in two ways: $13|24$ and $14|24$. 
Choosing a leaf from each group $1\ldots4$ gives a quartet \textit{around}  branch $Q$ (e.g., yellow dashed). 
(c) Visualizing  quartet support using DiscoVista. 
Bars shows quartet frequencies around two internal branches (labelled 6 and 7) of the tree shown in orange.  
Red bars correspond to the frequency of the main resolution shown in the tree and blue bars are for alternative resolutions, identified with branch lables matching the tree. 
(d)
Accuracy of ASTRAL branch length in simulations \citep{localpp-sm} on true and estimated gene trees (alignments length: 250bp to 1500bp). 
(e)
Accuracy of localPP in the same simulations.
For a threshold $p$,
branches of the ASTRAL tree and the two rearrangements around each of them ($3(n-3)$ in total) are categorized into 
true positives (correct, support $\geq p$), 
true negatives (correct, support $< p$),
false positives (incorrect, support $\geq p$),
false negatives (incorrect, support $< p$), and recall and false positive rates are computed.
Exploring $p$ produces the ROC curve where a higher line means more true positives for each false positive rate.
LocalPP dominates MLBS. 
(f) Testing the polytomy null hypothesis for 7 recalcitrant branches (A-F) in the ASTRAL species tree computed from 2022 supergene trees of an avian dataset. 
With more genes, p-values drop for some branches but not for others \citep{polytomy}. 
Branch E (position of enigmatic species Hoatzin) seems to be best explained by a hard polytomy. 
}
\label{fig:quartet}
\end{figure}

ASTRAL estimates coalescent units (CU) lengths of all internal branches and  of terminal branches corresponding to species with multiple individuals. 
True CU branch lengths are proportional to the number of generations spanned by the branch and inversely proportional to the population size \citep{Degnan2009}.
CU is important in MSC modeling because branch CU length  is what identifies the amount of topological discordance. 
Shorter branches lead to more discordance, especially when adjacent to each other. 
For a quartet, if the length of the only internal branch is $d$ in CU, the probability of a gene tree matching the species tree is $1-\frac{2}{3}e^{-d}$ and the probability of each of the two alternative topologies is $\frac{1}{3}e^{-d}$ (Fig~\ref{fig:quartet}a).
Thus, when a quartet gene tree appears $f>\frac{1}{3}$ times, $-\ln \frac{3}{2} (1-f)$ gives an estimate of the branch length.

ASTRAL exploits this observation to compute CU branch lengths (with simplifying assumptions) using a fast algorithm for computing mean quartet frequencies ``around'' each branch (Fig~\ref{fig:quartet}b). 
\cite{localpp-sm} have shown that despite assumptions, ASTRAL CU branch lengths are accurate {\em when} a sufficiently large number of true gene trees are used (Fig~\ref{fig:quartet}d). 
ASTRAL CU branch lengths, however, suffer from two issues, which limit their usability in practice. 
An obvious  shortcoming is that terminal branches for single-individual species lack an estimated length, limiting the utility of the computed branch lengths.

The second difficulty is the lack of robustness to gene tree error.
Gene tree error tends to increase gene tree discordance; as ASTRAL branch lengths are only a function of discordance (and nothing else), gene tree error results in under-estimation of branch lengths. 
For example, 
\cite{localpp-sm} observed lengths that were close to an order of magnitude underestimated for the least strong gene trees they tested (Fig~\ref{fig:quartet}d). 
In conditions where gene trees had even moderately high resolution (e.g., $60$\% mean BS corresponding to 1500bp genes), estimated  lengths were relatively accurate.

\subsection{Branch Support using Local Posterior Probability (localPP)}
The traditional method for obtaining branch support for species trees is multi-locus bootstrapping (MLBS) \citep{Seo2008}.
MLBS first bootstraps gene trees and then runs the summary method in replicate runs using bootstrapped gene trees as input. 
In the end, support values are computed by counting how often a branch appears in this collection of bootstrapped species trees.  
The MLBS method has turned out to have severe limitations \citep[e.g.,][]{Simmons2019}. 
For example, \cite{mrl-sysbio} showed in simulations that MLBS tends to both overestimate and underestimate support. 
The heart of the problem is the increased discordance among bootstrapped gene trees, compared to ML gene trees (which themselves tend to overestimate conflict). 
Recall that each locus can be relatively short and lacking in informative sites, a condition that is not conducive to accurate bootstrapping \citep{bootstrap}. 
To address limitations of MLBS, \cite{localpp-sm} designed a way to compute branch support for ASTRAL trees, without bootstrapping. 

If a branch in an estimated species tree is correct, for every quartet selected around it, the probability of observing the species tree should be at least $\sfrac{1}{3}$. 
Thus, asking whether a branch is correct is akin to asking whether the true probability of all quartet trees around the branch appearing in gene trees is higher than $\sfrac{1}{3}$. 
Given the distribution of quartet frequencies in an error-free sample of gene trees,  Bayes's rule can be used to compute the probability that the probability of observing the quartet tree is above $\sfrac{1}{3}$. 
This probability will give use the posterior probability (PP) of the branch being correct, given the input gene trees. 
For four species, the PP can be computed analytically (with a choice of a convenient prior distribution on branch lengths, which corresponds to assuming the species tree is generated under the Yule model). 

Exact calculation of PP is difficult for more than four species. 
However, with several simplifying assumptions, we can fall back to the case of four species.  A main assumption is \textit{locality}: in computing the support for a branch, we assume that all four branches around it are correct, enabling us to only consider three rearrangements around the branch (Fig.~\ref{fig:quartet}b). 
Because of this assumption, this measure of support is called localPP. 
\cite{localpp-sm} show in simulations that localPP is more accurate than MLBS when gene trees are inferred using short loci (i.e., from gene trees with relatively high error) and matches MLBS when gene trees are highly accurate (Fig.~\ref{fig:quartet}e). 
Moreover, localPP does not require bootstrapping gene trees and therefore is much faster than MLBS. 
Since its introduction, localPP has been adopted by many studies.

\section{Followup Analyses and Visualization}\label{sec:followup}
Several analyses can follow the species tree inference to gain additional insights. 
These followup analyses can be performed on any tree, whether computed by ASTRAL or not. 
ASTRAL can perform the following analyses and  compute localPP for any tree given to it using the {\tt -q} option.

\subsection{Testing for Polytomies}
A central question in systematics is whether a particular branch is resolvable given the present data or more broadly, at all.
Branches that cannot be resolved are removed, resulting in polytomies. 
Polytomies are called \textit{hard}  when the multifurcation is biological, and no amount of data should be able to resolve it,  or \textit{soft} when the present data cannot resolve the relationships due to lack of power. 
\cite{polytomy} introduced a  frequentist approach for testing the following null hypothesis: a given branch in the tree has length zero and should be contracted.
Note that under the null hypothesis, the quartet frequencies around a branch are $\sfrac{1}{3}$ for all three resolutions around the branch. A simple Chi-squared test can be used to test this null hypothesis. 
However, the failure to reject the null is not the acceptance of null; thus, when the null hypothesis is not rejected, we replace the branch with a polytomy, but we cannot say if it is a soft or a hard polytomy. 
\cite{polytomy} showed in simulations that the method successfully controls the false positive rate and is powerful in rejecting the null, given sufficient genes. 
The method also showed intriguing patterns when applied to the base on Neoaves (Fig.~\ref{fig:quartet}f), indicating that some (but not all) recalcitrant branches should perhaps be replaced with polytomies. The test for polytomies is implemented in ASTRAL and can be invoked using the option {\tt -t 12} (see ASTRAL documentations).

\subsection{Per Branch Quartet Support (Measure of Discordance)}

Phylogenomic studies are often interested in the amount of discordance {\em per branch} of the species tree. 
The computation of branch length and localPP in ASTRAL is contingent on first computing, for each branch, its quartet support.
Note that around each branch of an unrooted tree, many quartets are defined (Fig.~\ref{fig:quartet}b) that map to that branch and only that branch (there are $n-3$ to $(\sfrac{n}{4})^4$ such quartets per branch).

The quartet support of a species tree branch with respect to gene trees (with no missing leaves) is the proportion of times that quartets around the branch are  resolved identically to the species tree in the gene trees. 
When there are missing data, the definition becomes more tricky because several normalization schemes become possible. 
We use the following definition.
First, we discard all genes that do not fully include {\em any} of the quartets around the branch. 
Then, for each gene, we compute what portion of its quartets supports each topology, and we compute the mean of these values over all genes.
Thus, we get a number between 0 and 1 for each quartet topology around the branch. 

The quartet score of a branch can be used as a measure of discordance around the branch. 
ASTRAL can output quartet scores for all branches of a given tree ({\tt -t 1} and {\tt -t 8}). Several points are worth mentioning:
\begin{itemize}
    \item Values close to $\sfrac{1}{3}$ point to very high levels of discordance. 
    However, given a large number of gene trees, quartet scores that have relatively small divergences from $\sfrac{1}{3}$ (e.g., 40\%) can lead to high localPP.
    Also, remember that discordance includes both true discordance and the effects of gene tree error. 
    \item Under ILS, one expects quartet scores of the second and third topologies to be identical. {When the two frequencies diverge substantially, ILS assumptions are violated, either during gene tree estimation  (e.g., due long branch attraction) or because other biological sources of discordance (e.g., paralogy) also exist. Both cases warrant extra caution in interpretation.}
    \item In rare occasions, a branch of the ASTRAL tree has a quartet score below $\sfrac{1}{3}$. This can happen for several reasons, but in all cases, the branch should be considered unresolved (will have a localPP of 0). 
\end{itemize}

\paragraph{DisvoVista.}
The best way to interrogate the quartet scores produced by ASTRAL is to visualize them for branches of interest. 
\cite{discovista} have developed a tool called DiscoVista to visualize quartet scores around important branches (and also produce other visualizations of discordance). 
For example, in Figure~\ref{fig:quartet}c, we summarize quartet scores around two focal branches of the plant tree from \cite{1kp-pilot}; helpfully, DiscoVista collapses large groups into individual nodes for better visualization.



\section{Conclusion}\label{sec:conc}
I reviewed the relatively substantial body of knowledge available in the literature on the ASTRAL method, including best practices for using it. 
I hope the reader comes away with these messages:
\begin{itemize}
    \item ASTRAL is a statistically consistent method of species tree estimation given inputs sampled with no error under the MSC model.
More generally, it is consistent under any model for which the quartet that matches the species tree is expected to occur with the highest frequency.
    \item ASTRAL is extremely scalable and can analyze many thousands of species. 
    \item ASTRAL, like other summary methods, can be sensitive to gene tree estimation error, a problem that is alleviated but not eliminated if {\em extremely} low support branches in gene trees are contracted.
    \item ASTRAL has performed well in terms of accuracy in simulation analyses compared to other summary methods. The performance with respect to  concatenation depends on the amount of discordance and phylogenetic signal in input genes. 
    \item ASTRAL's  native localPP is a better method of computing support than multi-locus bootstrapping. 
    \item On real data, care is needed for preparing the input to ASTRAL, in particular, to avoid negative impacts of fragmentary data. However, extensive gene tree filtering is not recommended.
    \item ASTRAL is statistically inconsistent under models of gene evolution that include gene flow. 
    However, it has shown high accuracy under simulations with high levels of (randomly distributed) HGT.
\end{itemize}

\bibliographystyle{apalike}
\bibliography{references}
\end{document}